# Design of Fully Integrated 45 nm CMOS System-on-Chip Receiver for Readout of Transmon Qubit

Ahmad Salmanogli and Amine Bermak
Ahmadsalmanogli@gmail.com

*Abstract*— This study unveils a comprehensive design strategy, intricately addressing the realization of transmon qubits, the design of Josephson parametric amplifiers, and the development of an innovative fully integrated receiver dedicated to sensing ultra-low-level quantum signals. Quantum theory takes center stage, leveraging the Lindblad master and quantum Langevin equations to design the transmon qubit and Josephson parametric amplifier as open quantum systems. The mentioned quantum devices engineering integrated with the design of a fully integrated 45 nm CMOS system-on-chip receiver, weaves together a nuanced tapestry of quantum and classical elements.
On one hand, for the transmon qubit and parametric amplifier operating at 10 mK, critical quantum metrics including entanglement, Stoke projector probabilities, and parametric amplifier gain are calculated. On the other hand, the resulting receiver is a symphony of high-performance elements, featuring a wide-band low-noise amplifier with a 0.8 dB noise figure and ~37 dB gains, a sweepable 5.0 GHz sinusoidal wave generator via the voltage-controlled oscillator, and a purpose-designed mixer achieving C-band to zero-IF conversion. Intermediate frequency amplifier, with a flat gain of around 26 dB, and their low-pass filters, generate a pure sinusoidal wave at zero-IF, ready for subsequent processing at room temperature. This design achieves an impressive balance, with low power consumption (~122 mW), a noise figure of ~0.9 dB, high gain (~130 dB), a wide bandwidth of 3.6 GHz, and compact dimensions (0.54*0.4 mm²). The fully integrated receiver capability to read out at least 90 qubits positions this design for potential applications in quantum computing. Validation through post-simulations at room temperature underscores the promising and innovative nature of this design.

## I. Introduction

Transmon qubits play a critical role in quantum computing, known for their long coherence times and reduced sensitivity to charge noise [1-7]. The transmon qubit is a type of superconducting qubit that has gained prominence in the field of quantum computing due to its improved coherence properties. It is essentially an artificial atom created using Josephson junctions, which are superconducting devices that exhibit non-linear current-phase relationships. The transmon qubit is designed to operate in the regime where its energy levels are anharmonic, meaning that the energy separation between its quantum states is not constant. This anharmonicity helps mitigate the negative effects of charge noise, enhancing the qubit's coherence time and making it more robust against errors [1-3]. Unlike the traditional charge qubits, the transmon qubit operates with a large junction capacitance, resulting in a reduced sensitivity to charge fluctuations. This design choice, combined with the deliberate engineering of an anharmonic spectrum, contributes to the transmon qubit's improved coherence and ease of control. Transmon qubits find applications in quantum information processing tasks, particularly in the development of quantum computers. Their enhanced coherence properties make them suitable for implementing quantum gates and performing quantum algorithms with reduced error rates. The robustness of transmon qubits against certain types of noise positions them as promising candidates for the realization of scalable and fault-tolerant quantum computing architectures [7-9]. The transmon qubit, with its unique design and improved coherence characteristics, exhibits remarkable entanglement and superposition properties that render it particularly well-suited for various quantum applications. Entanglement properties of transmon qubits [4, 8] can be leveraged to achieve quantum-enhanced sensing capabilities, surpassing classical limits in measurement precision. This has implications for fields like quantum-enhanced imaging and sensing technologies. In addition, the entanglement and superposition properties of transmon qubits position them as key components in the advancement of quantum technologies. The fully integrated receiver [10-21] in quantum computing play a critical role in extracting information from transmon qubits. These circuits are essential for measuring the quantum state of a qubit accurately. There are some key considerations and features that designers should pay attention to when developing system-on-chip (SoC) receiver for transmon qubits: 1) measurement process not to disturb the quantum state of the qubit during or after the measurement; 2) High fidelity in measurement to minimizing errors in the measurement process; 3) Fast readout speed; 4) Crosstalk minimizing, the readout of one qubit does not interfere with the states of neighboring qubits. 5) Scalability, the designed receiver must be scalable to accommodate a growing number of qubits; 6) Receiver should be designed to be robust against various sources of noise, including thermal noise and environmental fluctuations. Filtering techniques can be employed to enhance signal-to-noise ratios and improve measurement accuracy; 7) Readout circuits must interface effectively with classical control electronics and signal processing systems for comprehensive quantum information processing; 8) The SoC receiver is positioned close to the qubit to minimize signal losses. In summary, designed circuits as readout for transmon qubits [12, 19-21] need to balance factors such as speed, fidelity, scalability, and noise resilience. The state-of-the-art involves sophisticated techniques that leverage advances in superconducting circuits and amplification technologies to achieve high-performance quantum measurements.

This study, on one side, focuses on advancing the design of a transmon qubit [1-6], employing quantum theory to design transmon qubit [1-3] and Josephson parametric amplifier (JPA) [22-24]. On the other side, the fully integrated 45 nm CMOS SoC receiver as a readout of the transmon qubits is presented. Transitioning to the cryogenic phase, the circuit employs advanced 45 nm CMOS technology for components like low-noise amplifiers (LNA), voltage-controlled oscillators (VCO), mixers, Intermediate frequency (IF) amplifier and filters. Notably, the wide-band LNA achieves a noise figure of 0.8 dB and a gain of 37 dB. The resulting output is a zero-IF pure sinusoidal wave, optimized for room-temperature processing. Emphasizing low power consumption, minimal noise interference, high gain, and compact dimensions, this study pioneers an active element-centric design paradigm, verified through comprehensive simulations at room temperature.

## II. DESIGN OF TRANSMON QUBIT

The schematic of the system is illustrated in Fig. 1 (in the middle section), in which two qubits, are embedded into a TL. It is supposed that, a TL with a typical length is driven with a radio-frequency (RF) wave through coupling capacitor $C_{in}$. The generated wave across the TL is coupled capacitively, $C_g$, to the qubit. A capacitor ($C_d$) is attached to control the coupling energy of the transmon qubit. In a transmon qubit to create the ultra-coupling ($E_J \gg E_c$), the charging energy in qubit ($E_c$) is decreased rather than the direct increase of $E_J$ [1]. Accordingly, to increase the TL energy coupling to the transmon qubit, one can manipulate $C_g$ through which the coupling factor between the TL and transmon qubit varies. In the model presented, $C_J$ is qubit capacitance. Finally, the output of the amplified quantum signals is transferred through the TL and capacitively coupled ($C_{out}$) to the output's TL. In the following, the fully integrated SoC receiver circuit is designed and attached to sense, amplify, mix, and generate the suitable zero-IF of the quantum signals.

The standard method to describe the quantum behavior of an electric circuit includes finding the classical Hamiltonian of the model and then imposing the canonical commutation relation on its degree of freedom to generate the quantum Hamiltonian [25]. Regarding the model considered and illustrated in Fig. 1, the transmon qubit Hamiltonian is given by:

$$H_q = \sum_m \hbar \omega_m \left( a_m^+ a_m + \frac{1}{2} \right) \\ + \sum_{m,n} \left\{ \left( 2e\beta_m \sqrt{\frac{\hbar}{2\omega_m}} \right) (a_m - a_m^+) \frac{1}{\sqrt{2}} \sqrt[4]{\frac{E_J}{8E_c}} (b_n - b_n^+) \right\} \quad (1) \\ + \sum_{m,n} \sqrt{8E_{cm}E_J} \left( b_n^+ b_n + \frac{1}{2} \right) - \frac{E_{cm}}{12} (b_n + b_n^+)^4$$

where $\hbar$, $\omega_m$, $E_J$, $e$, subscript m, and n are the reduced Planck constant, TL harmonic frequency, Josephson junction energy, electron charge, TL mode number, and qubit mode number, respectively. To express the quantum Hamiltonian in terms of the ladder operators, the "cos" function in the transmon qubit Hamiltonian is expanded in the form of the Taylor series. In above equation, ($a_m^+$, $a_m$) and ($b_n^+$, $b_n$) are the creation and annihilation operators of the TL and transmon qubit modes, respectively. Additionally, the remaining parameters are defined as $\beta_m = \alpha_m \omega_m/(1-\sum_m \alpha_m^2)\sqrt{(C_G)}$, $\gamma_m = (e\beta_m/\hbar)(E_J/8E_{c0})^{0.25}\sqrt{(\hbar/\omega_m)}$, and $E_{cm} = E_c/(1-\sum_m \alpha_m^2)$, where $E_{c0} = e^2/2C_G$ is the transmon qubit charging energy. Furthermore, the following definitions are made as $C_G = C_g + C_J + C_B$ and $\alpha_m = C_g u_m(X_J)/\sqrt{(C_\Sigma C_G)}$ [4].

In the following, it is necessary to apply the Lindblad master equation on the total Hamiltonian of the quantum system interacting with the environment. The total Hamiltonian is defined as $H_t = H_q + H_{in} + H_{out}$, where $H_{in}$ and $H_{out}$ are input and output Hamiltonians due to the capacitively coupling of the $V_g$ as a voltage control and capacitively coupling of the transmon qubit to the output circuit, respectively. The Lindblad master equation [25-26] is applied to derive the quantum system's dynamics equation of motion interacting with its environment. In this article, similar to the literature [25, 27], it is supposed that the system is expanded to include the environment; thus, the expanded system (the TL with qubit implemented is combined with the environment) can be considered a closed quantum system. A closed quantum system evolution is governed by the von Neumann equation derived as [26-28]:

$$\dot{\rho}(t) = \frac{1}{j\hbar}\left[H_t, \rho_t(t)\right] \quad (2)$$

where $\rho(t)$ and $H_t$ are the density matrix describing the probability distribution of the quantum state and system total Hamiltonian expressed above. To solve the Lindblad master equation, the Qutip toolbox in Python [26] is utilized to determine the time evolution of the system's density matrix and ultimately, the covariance matrix (CM) for the quadrature operators $X_i$ and $Y_i$, where i = 1, 2. These operators are defined for the quantum system using the formulae $X_i = (a_i + a_i^+)/\sqrt{2}$ and $Y_i = (a_i - a_i^+)/j\sqrt{2}$. In Fig. 1a, the graph depicts the photon average numbers of the transmon qubit (depicted in red), TL (in black), and cross-correlation photon numbers (in blue). This representation reveals a mixing behavior arising from the coupling between the TL and the qubit. The entanglement between modes is calculated using the illustrated average photon numbers, and the outcome is presented in Fig. 1b. It is evident from Fig. 1a that, for the establishment of entanglement, the amplitude of cross-correlation photon numbers needs to be increased compared to the qubit and TL modes. For entanglement analysis, the entanglement metric is calculated using the absolute value of the cross-correlation between the qubit and TL modes, divided by the square root of the product of the qubit and TL photon numbers [27-28]. The entanglement metric greater than unity means that the two modes become entangled. A value close to 0 indicates weak or no entanglement, while a value close to unity or greater than it, indicates strong entanglement between the qubit and TL modes. The augmentation of cross-correlation photon numbers directly contributes to the coupling between TL and qubit modes, thereby significantly manipulating the entanglement between modes. For enhanced clarity, a vertical blue-dashed line is incorporated into the graph to delineate the limit of entanglement. Significantly, the study underscores that both the amplitude of photons and the entanglement between modes exhibit strong dependence on factors such as $C_d$, $C_j$, $C_g$, input and output coupling capacitors, as well as the TL's contributed capacitance and inductance.

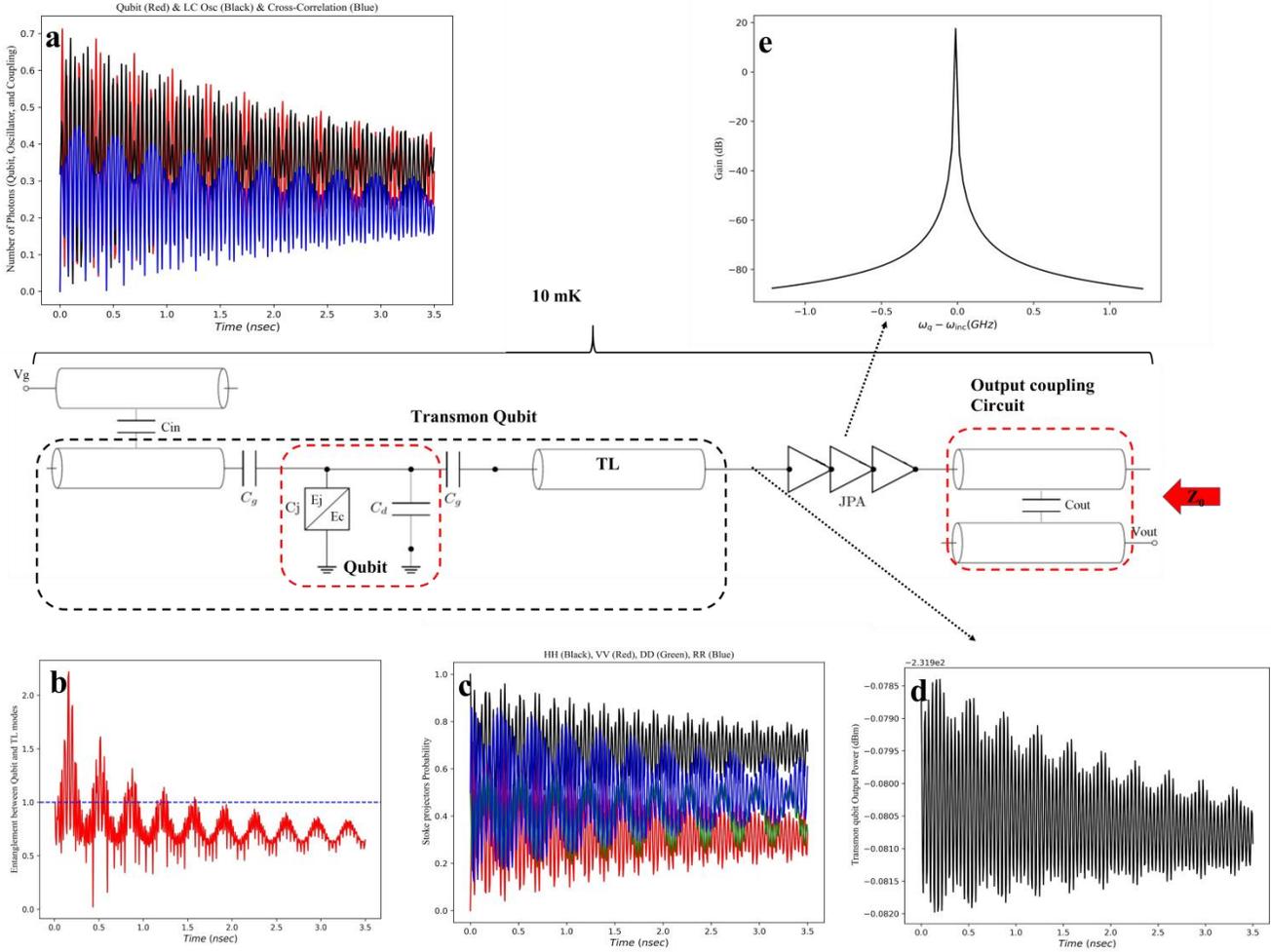

Fig. 1. Schematic of the proposed structure containing qubit coupling to TL through $C_g$ (in the middle of the graph) followed by a series of JPA to amplify the quantum signal, and finally coupling to the readout circuit to transfer the signal; $C_g = 20$ fF, $C_d = 40$ fF, $C_J = 2$ fF, $C_0 = 0.66$ pF/m, $L_0 = 623$ nH/m [2]; a) the qubit and TL average number of photons, b) entanglement between modes using entanglement metric, c) Stoke projection probability ($P_{HH}$, $P_{VV}$, $P_{DD}$, and $P_{RR}$), d) output signal power, e) the gain of the single JPA.

In quantum mechanics, the Stokes and Pauli probabilities are calculated to characterize the state of a quantum system after a measurement [25, 29-30]. These probabilities provide information about the system's behavior, particularly with respect to the measurement outcomes corresponding to different measurement bases. The Stokes parameters (depicted in Fig. 1c) can provide a way to describe the polarization state of mode. For a two-level quantum system (qubit in this design), the Stokes parameters are related to the probabilities of measuring the qubit in different states after a polarization measurement. The calculation of Stokes probabilities allows to understand the correlations between different measurement outcomes. This information is critical for determining how the quantum system responds to measurements and how measurements in one basis may affect measurements in another basis. For entangled quantum systems (qubit and TL modes), the Stokes probabilities provide a means to study correlations between the measured outcomes of different subsystems. The entanglement properties of a quantum state can be revealed through the analysis of probabilities associated with various measurement bases. To analyze entanglement using Stokes parameters, one can compute the concurrence ($2|P_{DD}*P_{RR} - P_{HH}*P_{VV}|$), which is a measure of entanglement. The concurrence results is shown in Fig. 2. The concurrence ranges between 0 (no entanglement) and 1 (maximal entanglement). If the amplitude of the concurrence is changing between zero and 0.5, it means that the system is exhibiting varying degrees of entanglement over time. A concurrence close to 0 indicates a separable state (no entanglement), and A concurrence close to 1 indicates a maximally entangled state.

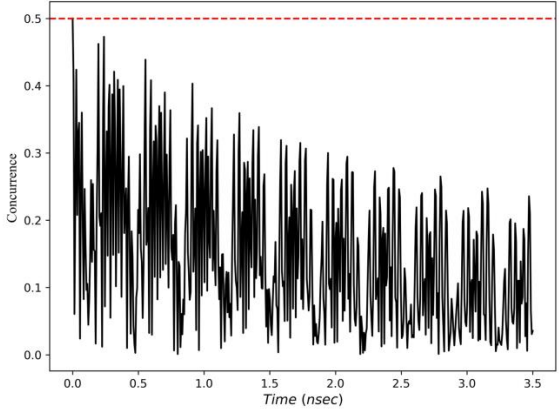

Fig. 2. Concurrence between qubit and TL modes

The discrepancy between the entanglement metric and concurrence results may arise from the fact that these two measures capture different aspects of entanglement and are sensitive to different aspects of the quantum state. The entanglement metric applied is a measure of correlation between the qubit and TL modes [4, 8], normalized by the square root of the product of their individual photon numbers. This metric can provide insight into how the modes are correlated and entangled during the evolution. On the other hand, concurrence is a measure specifically designed for quantifying the entanglement between two qubits. It is defined based on the density matrix of the two-qubit system and involves the calculation of eigenvalues.

The present investigation reveals that the signal produced by the transmon qubit is approximately -220 dBm, making direct transmission unfeasible due to the extremely low amplitude of quantum signals. To address this, signal amplification becomes essential at a sub-cryogenic temperature of around 10 mK. The primary candidate for such an operation, potentially the sole one, is the Josephson parametric amplifier (JPA) [22-24]. Although some studies propose electronic-based circuits utilizing InP HEMT [31-33] and CMOS [17-18, 24], and their nonlinearity for quantum signal amplification, noise figure, even under ideal conditions, is incomparable to JPA. A JPA is a device based on Josephson junctions and operates in the quantum regime [22-24]. It is used to amplify signals in quantum circuits, particularly in the field of quantum information processing and quantum communication. JPAs have the capability to provide quantum-limited amplification, meaning they can amplify weak quantum signals while adding a minimal amount of noise to the output. This property is crucial for preserving the quantum information encoded in the signals. JPAs can be employed in quantum systems for non-destructive readout of quantum states. This is particularly useful in quantum computing, where preserving the state of a qubit during readout is essential. JPAs play a role in quantum signal processing tasks, such as quantum state tomography [34-35], where the amplification of signals is required for accurate measurement and characterization of quantum states. State-of-the-art JPAs are often integrated into superconducting qubit systems. This integration is essential for tasks such as qubit readout and the implementation of quantum error correction protocols. Ongoing research focuses on improving the gain and efficiency of JPAs to enhance their applicability in quantum information processing. This includes optimizing the design of Josephson junctions and resonators. Researchers are exploring techniques to implement noise squeezing in JPAs, which can further reduce the added noise during signal amplification. Noise squeezing is a quantum phenomenon that allows for noise reduction in one parameter at the expense of increased noise in another [36-39]. Consequently, JPAs play a crucial role in quantum circuits, providing quantum-limited amplification and enabling various applications in quantum information processing.

In the study's designed transmon qubit, it is demonstrated that either through the nonlinearity terms in the total Hamiltonian, or via coupling with TL modes, signal amplification can occur. To show this point, we theoretically derive the quantum Langevin equation [25, 40-44] using the total Hamiltonian of the system and after linearization around the cavities strong field [40-44] the simplified equations are given by:

$$\dot{\delta a_m} = -(i\omega_m + \frac{\kappa_1}{2})\delta a_m + i\beta'(\delta b_n + \delta b_n^+) - i2\alpha_{out}(\delta a_m + \delta a_m^+) + \sqrt{2\kappa_1}\delta a_{in}$$

$$\dot{\delta a_m^+} = -(-i\omega_m + \frac{\kappa_1}{2})\delta a_m - i\beta'(\delta b_n + \delta b_n^+) + i2\alpha_{out}(\delta a_m + \delta a_m^+) + \sqrt{2\kappa_1}\delta a_{in}^+$$

$$\dot{\delta b_n} = -(i\omega_q + \frac{\kappa_2}{2})\delta b_n + i\beta'(\delta a_m + \delta a_m^+) + i2\omega_k |\beta_{con}|^2 \delta b_n^+ + \sqrt{2\kappa_2}\delta b_{in}$$

$$\dot{\delta b_n^+} = -(-i\omega_q + \frac{\kappa_2}{2})\delta b_n - i\beta'(\delta a_m + \delta a_m^+) - i2\omega_k |\beta_{con}|^2 \delta b_n^+ + \sqrt{2\kappa_2}\delta b_{in}^+$$

(3)

In this equation $\alpha_{out} = C_{out}/[2Z_{out}*(C_{out}C_\Sigma + C_\Sigma C_0 l_{TL} + C_{out}C_0 l_{TL})]$, where $C_{out}$ is the coupling capacitance of the transmon qubit to the output TL, $Z_{out}$ is the output TL characteristics impedance, $l_{TL}$ is the length of TL, and $C_0$ is the TL capacitance. In addition, $\beta' = (2e/\hbar)(\beta_m/\sqrt{2})(E_j/8E_c)^{0.25}\sqrt{(\hbar/2\omega_m)}$, $\hbar\omega_k = E_c$, which is the JJ charging energy, and finally $|\beta_{con}|$ is the qubit average number of photons. By transferring into the Fourier domain and with the assumption that the $\kappa_1 = \kappa_2 = \kappa$, the intra-cavity modes ($\delta a_m$ and $\delta b_n$) can be expressed in a matrix form in terms of the input modes as:

$$\begin{bmatrix} \delta a_m \\ \delta a_m^+ \\ \delta b_n \\ \delta b_n^+ \end{bmatrix} = \begin{bmatrix} i\delta\omega_1 + i2\alpha_{out} + \frac{\kappa_1}{2} & i2\alpha_{out} & -i\beta' & -i\beta' \\ -i2\alpha_{out} & -i\delta\omega_1 - i2\alpha_{out} + \frac{\kappa_1}{2} & i\beta' & i\beta' \\ -i\beta' & -i\beta' & i\delta\omega_2 + \frac{\kappa_2}{2} & -i2\omega_k|\beta_{con}|^2 \\ i\beta' & i\beta' & i2\omega_k|\beta_{con}|^2 & -i\delta\omega_2 + \frac{\kappa_2}{2} \end{bmatrix}^{(-1)} \sqrt{2\kappa} \begin{bmatrix} \delta a_{in} \\ \delta a_{in}^+ \\ \delta b_{in} \\ \delta b_{in}^+ \end{bmatrix} \quad (4)$$

where $\delta\omega_1 = \omega_m - \omega_{inc}$ and $\delta\omega_2 = \omega_q - \omega_{inc}$ are the detuning frequency, and $\omega_{inc}$ is the incident angular frequency. Using the inverse of the scattering matrix expressed in Eq. 4, and the input-output formula [25-27], one can calculate the gain of the quantum system designed. Notice that the blue-dashed, red-dashed, and green-dashed rectangles indicate the effect of the output TL coupling, coupling of the transmon qubit and TL modes to each other, and nonlinearity of the qubit, respectively. The output TL can change the diagonal elements, while the transmon qubit modes coupling to TL and the nonlinearity of the transmon qubit just change the off-diagonal elements in the scattering matrix. The simulation results are depicted in Fig. 1e. It is demonstrated that, at a detuning frequency near zero, the designed transmon qubit functions as a JPA. This occurs due to parametric features arising from the nonlinearity of the inductor or the coupling between the qubit and TL modes. The amplification gain is highly influenced by the qubit's position, qubit and TL modes coupling, and the average photon number of the qubit and TL. Consequently, the parametric amplifier significantly enhances the quantum signal level, achieving up to -160 dBm.

Subsequently, the study shifts its focus to designing the analog circuit (receiver). This circuit, leveraging LNA and RF amplifier, detects and amplifies the mentioned quantum signals. It then utilizes a mixer to convert signals to zero-IF, followed by amplification and low-pass filtering using a series of IF amplifiers. This prepares the signals for digitization and subsequent processing at room temperature.

## III. DESIGN OF FULLY INTEGRATED SoC RECEIVER LEVERAGING 45NM CMOS

Designing the fully integrated 45 nm CMOS SoC receiver as a readout circuit for quantum computing applications requires careful consideration of various parameters to minimize the impact on quantum signals. The goal is to achieve high-fidelity measurements without introducing excess noise or distortion. The noise figure quantifies the amount of noise added by the receiver circuit. For quantum systems, low noise is critical to preserving the coherence of quantum states. The other important factor is the operational bandwidth. The receiver's bandwidth should match the frequency range of the quantum signals to capture all relevant information. Design filters, amplifiers, and mixers with appropriate bandwidths tailored to the quantum system's frequency range. Another important parameter that the designer should care about is the linearity. Nonlinearities in the analog circuit can distort quantum signals, affecting measurement accuracy. The other important factor is the integration of the receiver with quantum system; because quantum systems are sensitive, and the circuit should be designed to interface seamlessly without disturbing the quantum states. Consequently, a perfect readout circuit for quantum computing minimally affects quantum signals by prioritizing low noise, cryogenic compatibility, appropriate bandwidth, linearity, and efficient signal conversion. It requires a careful balance of these parameters to ensure high-fidelity measurements without compromising the delicate nature of quantum states [13-21].

Each component in the designed receiver including LNA, Mixers, VCO, and RF and IF filters plays a specific role in the readout process, contributing to the accuracy and efficiency of quantum state measurement. LNAs are crucial for amplifying weak signals from qubits during the readout process. As quantum states are inherently delicate, a low-noise amplification stage is essential to maintain the fidelity of the measured signal. Research is ongoing to develop LNAs with improved noise performance, wider bandwidths, and better integration with quantum systems [12-19].The post-processing outcomes of various subsystems within the analog readout circuit are presented below. In Fig. 3, the gain and noise figure (NF) of the designed LNA are depicted. The findings reveal that within the specified bandwidth (4.4-8.0 GHz), the NF of the designed circuit is consistently below 0.8 dB, and the gain exceeds 34 dB.

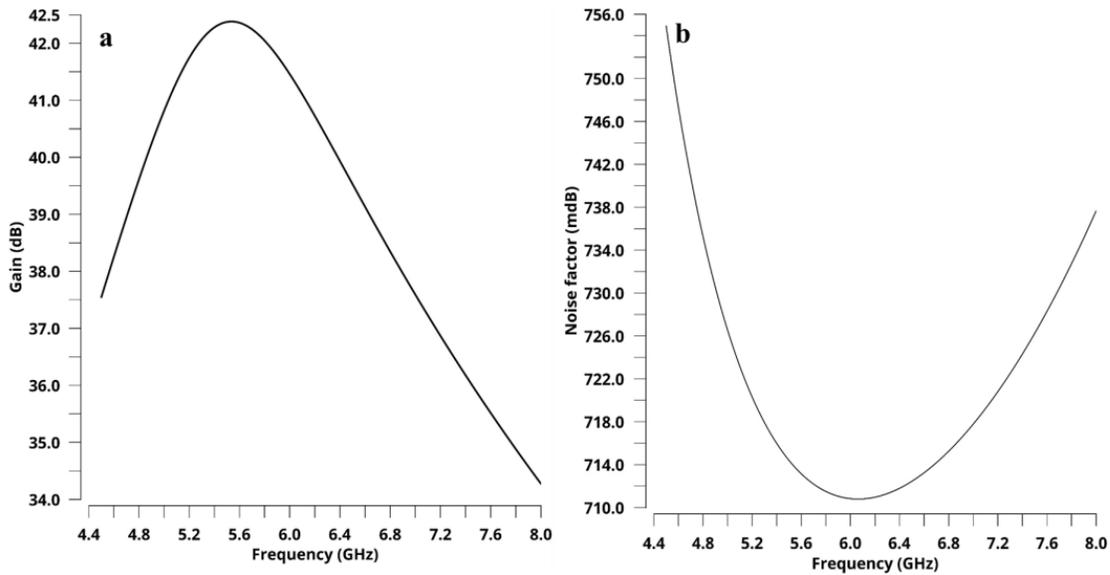

Fig. 3. LNA post-processing results; (a) gain vs. frequency, and (b) NF vs. frequency.

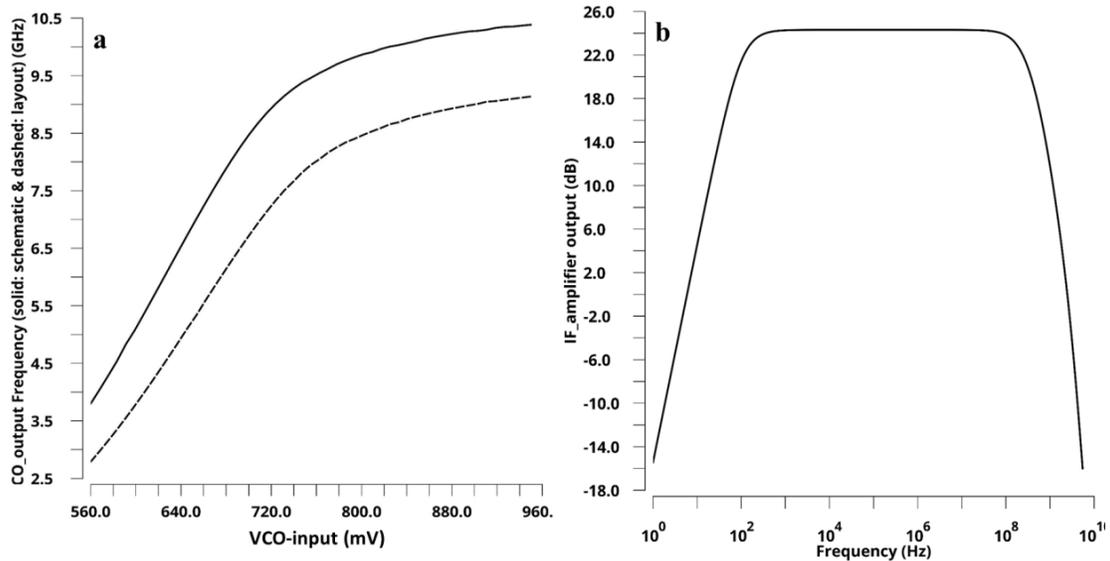

Fig. 4a) VCO output voltage signal vs. VCO input dc voltage signal (post-processing results); solid for the schematic and dashed for the post-processing results; b) IF post-processing results; gain vs. frequency.

Mixers are employed to down-convert the high-frequency qubit signal to an IF that is more suitable for further processing and measurement. Quantum-compatible mixers must be low-noise and operate at cryogenic temperatures. Recent developments focus on achieving higher linearity and bandwidth, essential for accurate qubit state discrimination.

VCOs generate a local oscillator (LO) signal that mixes with the qubit signal in the mixer stage. The LO signal is crucial for down-conversion. Research explores methods to enhance phase noise performance and frequency agility for diverse quantum applications. Fig. 4a displays the output of the VCO in both schematic and layout forms. The figure illustrates that by varying the VCO's input voltage from 650 mV to 960 mV, the output signal frequency changes from 2.5 to 10 GHz. As can be easily seen that there is a discrepancy between the schematic and the layout results; it may be due to some reasons such as 1) In layout, parasitic elements such as capacitance and resistance can significantly impact the performance of the circuit. These parasitic elements may not be accurately represented in the schematic, leading to differences in simulation results. 2) Layout simulation takes into account the actual physical layout of the components and interconnects. Signal integrity issues, such as transmission line effects, coupling, and crosstalk, may be more pronounced in the layout compared to the simplified representation in the schematic. 3) Simulations are often performed with corner cases and process variations to account for manufacturing tolerances. The impact of these variations may be more pronounced in layout simulations due to the physical proximity of components.

IF amplifier and filters are used to amplify and isolate the desired frequency components (zero-IF) and eliminate unwanted noise or interference. Compact and low-loss elements are critical for maintaining the coherence of quantum states. Fig. 4b showcases the output of the IF amplifier as the AC response, demonstrating its effectiveness as a band-pass filter for complete filtering of RF and very low frequencies.

The integration of LNAs, mixers, VCOs, and filters in circuit designed for quantum computing is essential to ensure the accurate and low-noise measurement of qubit states. Ongoing research and development in this field focus on improving the performance of these components at cryogenic temperatures, enhancing noise characteristics, and addressing the specific challenges posed by quantum systems. State-of-the-art components are tailored to the unique requirements of quantum circuits to enable progress in quantum computing technology [18-21]. The proposed SoC illustrated in Fig. 5 has been implemented in a standard 45nm CMOS process. It contains three RF amplifiers (LNAs), a VCO, a mixer, and four IFs. Noteworthy, the layout incorporates additional inset figures strategically placed to highlight specific sections for enhanced clarity and understanding. The comprehensive performance metrics of the circuit are noteworthy. The total power consumption of the circuit (overall circuit consumption) is measured at approximately 122 mW, underscoring its efficiency in power management. Impressively, the circuit demonstrates a substantial total gain of around 135 dB, attesting to its robust amplification capabilities. Furthermore, the circuit boasts a commendable noise figure, registering at approximately 0.9 dB, indicative of its ability to maintain signal integrity by minimizing unwanted noise contributions. These key parameters collectively underscore the circuit's prowess in amplification, power efficiency, and signal fidelity, making it well-suited for applications demanding high-performance characteristics. In the design of this circuit, the primary emphasis is placed on achieving a compact layout for the readout chip. The strategy employed involves a significant reduction in the use of inductors throughout the design. As evident in the layout, specific sub-devices, including the VCO and IF amplifier, have been designed without the incorporation of inductors. This deliberate choice contributes to the compactness of the overall chip. A noteworthy advantage of adopting inductor-free designs for the VCO and IF amplifier is the notable reduction in their physical dimensions compared to components like the mixer and LNA. This not only serves the primary objective of compactness but also facilitates a more efficient use of space on the chip. Furthermore, the absence of inductors in the VCO and IF amplifier design has secondary benefits, such as minimizing electromagnetic interference (EMI) concerns and simplifying the overall fabrication process.

The step-by-step post-processing analysis of the receiver circuit results is illustrated in Fig. 6. This figure provides insights into the key stages of signal processing, demonstrating the effectiveness of the circuit design. The Fig. 6a portrays the input signals, denoted as "tone," characterized by an ultra-low power level of approximately -160 dBm. This emphasizes the challenging starting point for a sensitive receiver, dealing with signals at the brink of detectability. The Fig. 6b showcases the output signal of the VCO, generating dominant frequencies around 5.0 GHz. While the main frequency components are highlighted, it is noted that additional tones are produced. The focus is on subsequent efforts to attenuate unwanted signals. One of the important point is to Highlighting the impact of RF amplifiers and filters; the Fig. 6c demonstrates the role of these components in refining the signal quality. It visualizes how the circuit manipulates the incoming signals, preparing them for further stages of processing. In Fig. 6d, the mixer output is presented, where the 5.2 GHz and 5.0 GHz input signals are mixed, resulting in a zero-IF signal at 200 MHz. This step is critical for down-conversion, and it indicates the successful transformation of signals to a manageable frequency range. Fig. 6e illustrates the signals after passing through two IF amplifiers. It emphasizes the effects of the Intermediate Frequency and the role of the filters, showcasing a substantial increase in the amplitudes of the zero-IF signals. This highlights the amplification and signal conditioning achieved in this stage. Fig. 6h and Fig. 6f collectively depict the final output signal's tones and their evolution over time. Notably, the output of the receivermanifests as a pure sinusoidal wave with a frequency around 200 MHz. This indicates the successful extraction and amplification of the desired signal. By comparing the RF signals with the zero-IF signals in Fig. 6h, the transformation from an initial RF signal at -160 dBm to a zero-IF signal with an amplitude of 10 dBm is evident. This highlights the circuit's capability to effectively convert very weak RF signals into a stronger, manageable form for subsequent processing. The sinusoidal nature of the output waveform emphasizes the purity and stability achieved by the designed circuit. This is crucial for applications requiring precision and accuracy in signal detection. As an interesting results, Fig. 6 effectively communicates the intricate stages of signal processing in the readout circuit, showcasing its ability to handle extremely low-power input signals and produce a robust, amplified, and frequency-transformed output.

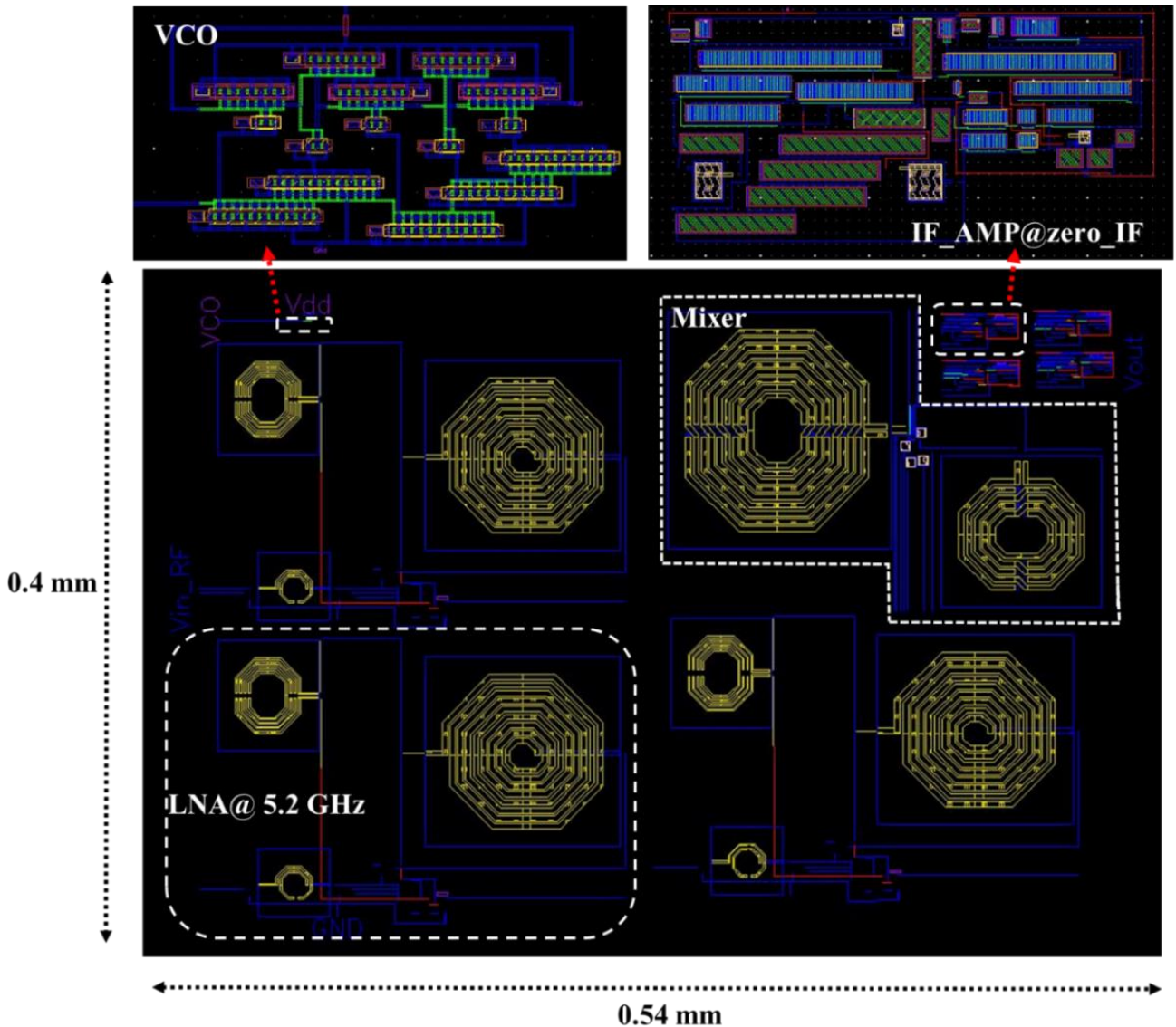

Fig. 5. The layout of the fully integrated SoC receiver containing the three RF amplifier (LNA), a VCO, a mixer, and four IF amplifiers; some inset figures are used to bold the a few unclear section of the layout.

Determining the number of qubits a fully integrated SoC receiver can support involves considering various factors related to the receiver's design and specifications. More crucial ones are the bandwidth of the receiver and the frequency range it covers. Qubits may operate at different frequencies, and the receiver should be able to accommodate the frequency range of the qubits. One can simply calculate the number of qubit supported by the receiver designed in this study using: Number of the qubits = (the bandwidth of the SoC Receiver)/(the require bandwidth for qubit). The overall circuit consumes 122mW with a 3.6 GHz bandwidth. If we suppose a very narrow but effective bandwidth by considering the LNA gain graph (Fig. 3), which is around 1.3 GHz. Considering a conservative 10MHz resonator bandwidth optimized for rapid qubit readout within time domains surpassing typical decoherence times and a 10MHz spacing between each qubit [19], our proposed analog circuit demonstrates the capability to readout up to 90 qubits within its bandwidth. This capacity not only enhances scalability but also results in an efficient power consumption of 1.35 mW per qubit. This compelling metric positions the designed system as an effective solution for large-scale quantum computing endeavors, underscoring both its performance and power efficiency in supporting a considerable number of qubits.

Table 1. Comparison table with state-of-the-art circuits as a receiver for qubit readout and the high-performance.

| | This work | Ref [19] | Ref [20] | Ref [21] |
|---|---|---|---|---|
| Operating Temperature | 300 K | 3.5 K | 300K | 3.5 K |
| Qubit platform | Transmon | Spin qubit/transmon | Si/SiGe Spin qubit | Transmon |
| System | Full-receiver | Full-receiver/PLL | Down-converter/VCO/Divider | Drive/full bias/readout |
| RF frequency | 4.4-8.0 GHz | 5-6.5 GHz | 240 GHz | 4.5-8 GHz |
| BW | 3.6 GHz | 1.4 GHz | 59 GHz | 3.5 GHz |
| Gain | 135 dB | 70 dB | 26 dB | 47 dB |
| NF | 0.9 dB | 0.55 dB | 24.5 dB | 1.1 dB |
| $P_{1dB}$ | >-45 dBm | >-85 dBm | >-27 dBm | >-49 dBm |
| VCO tuning range | 6 GHz | 6 GHz | 27 GHz | NA |
| Technology | 45 nm CMOS | 40 nm CMOS | 55 nm SiGe | 40 nm Bulk CMOS |
| Chip Area | 0.22 mm$^2$ | 2.8 mm$^2$ | 1.8 mm$^2$ | 5.2 mm$^2$ |
| Chip Power Consumption | 122 mW | 108 mW | 859 mW | 20 mW/channel |

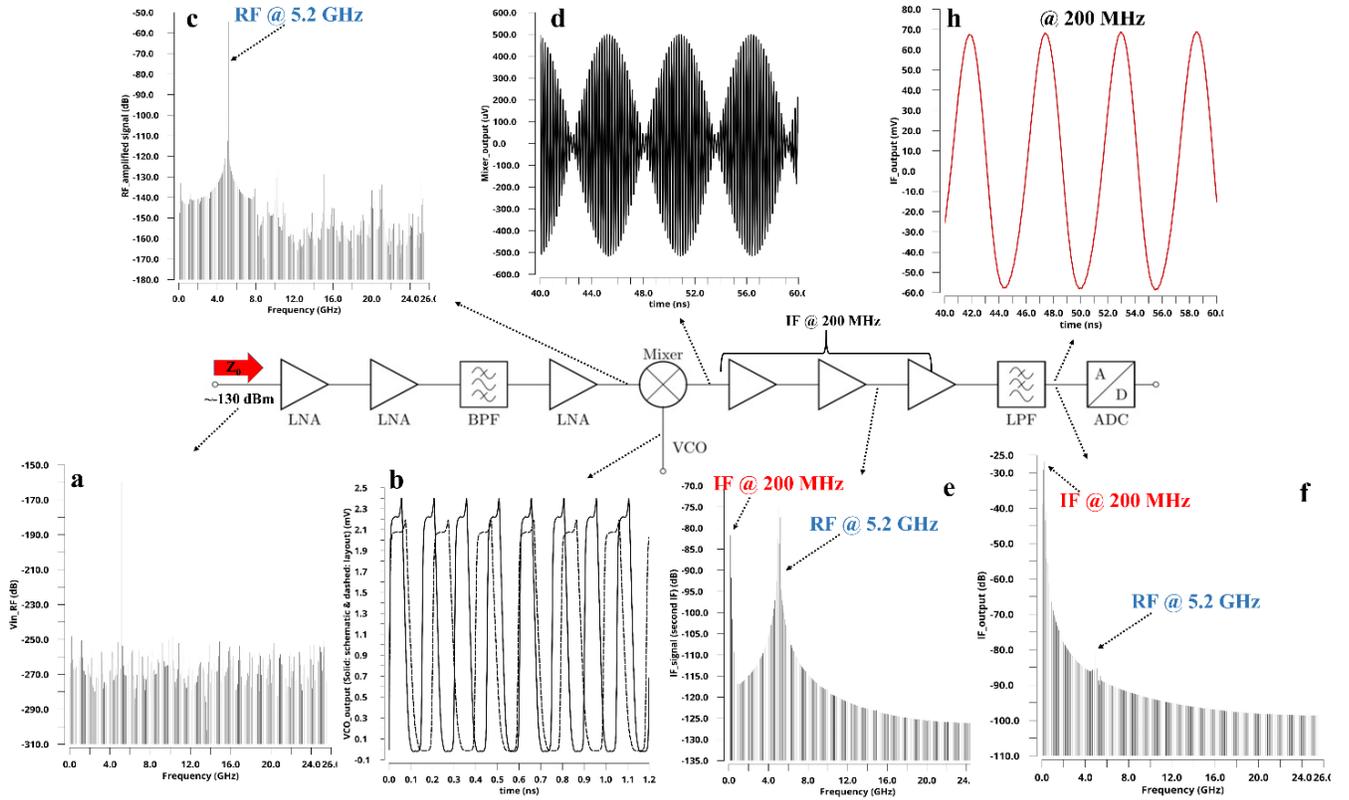

Fig. 6. Step-by-step of the post-processing results for the receiver designed; a) the spectrum of input signals tone, b) VCO output signal vs. time, c) the tones of the signal after RF amplification stage, d) mixer output vs. time, e) the signals spetrum after two IF amplifiers, f) and h) output signal tones and its time evolution.

IV. CONCLUSION

This research introduces an innovative quantum readout circuit, a fully integrated SoC receiver, tailored for transmon qubits, designed to function at an ultra-low temperature of 10 mK utilizing fully integrated 45 nm CMOS SoC receiver. A transmon qubit was meticulously subjected to a comprehensive analysis grounded in quantum theory. The qubit, coupled capacitively to a TL, was conceptualized as a quantum open system. In exploring this open quantum system, the Lindblad master equation was employed, unraveling key insights into crucial quantities such as the entanglement dynamics between the qubit and TL modes. Additionally, leveraging the total Hamiltonian of the system, Stoke projector probabilities, and transmon qubit output signals (expressed in output power in dBm) were calculated. The findings illuminated a significant aspect: the quantum signals generated by the transmon qubit were inherently low, necessitating amplification before integration into the analog circuit. Consequently, JPA was purposefully designed to serve as an effective amplifier operating at the challenging temperature of 10 mK. It was found that under zero detuning frequency conditions JPA proficiently enhance quantum signal levels around -160 dBm~-130 dBm. This transformative capability underscored the critical role of the JPA in preparing the quantum signals for subsequent stages of processing within the designed quantum readout circuit. The subsequent stages of the analog readout circuit were subjected to detailed design and analysis, resulting in exceptional performance metrics. The LNA demonstrated a remarkable noise figure below 0.8 dB, coupled with a gain exceeding 36 dB. The mixer efficiently generated a zero-IF signal, and the VCO showcased precise frequency control. Additionally, the IF amplifier effectively functioned as a band-pass filter, contributing to the circuit's overall signal conditioning. The comprehensive layout of the receiver circuit, comprising three RF amplifiers, a VCO, a mixer, and four IF amplifiers, underscores its compact and integrated design. Notably, the total power consumption of the fully integrated SoC receiver was constrained to approximately 122 mW, showcasing the circuit's efficiency. Impressively, it achieved a substantial gain of around 135 dB, accompanied by a noise figure of about 0.9 dB, highlighting the delicate balance between amplification and signal fidelity. The compact design and robust performance metrics, validated through simulations and analyses, position our circuit as a potential enabler for advancing quantum technologies. Crucially, the article addressed the application of the designed system by calculating its potential to support qubits. With an effective bandwidth of 1.3 GHz (the real bandwidth of the system is around 3.6 GHz), the design receiver exhibited the capability to read out up to 90 qubits, fostering scalability in quantum computing endeavors. The remarkable power efficiency, at 1.35 mW per qubit, positions the design as a compelling candidate for large-scale quantum applications. Building on these promising results, the quantum readout circuit emerges as a viable candidate for diverse applications in quantum computing, sensing, and medical imaging.

REFERENCES

[1] J. Bourassa, F. Beaudoin, Jay M. Gambetta, and A. Blais, "Josephson-junction-embedded transmission-line resonators: From Kerr medium to in-line transmon," Phys. Rev. A vol. 86, pp. 013814, 2012.

[2] Jens Koch, Terri M. Yu, Jay Gambetta, A. A. Houck, D. I. Schuster, J. Majer, Alexandre Blais, M. H. Devoret, S. M. Girvin, and R. J. Schoelkopf, Charge-insensitive qubit design derived from the Cooper pair box, Phys. Rev. A **76**, 042319 – Published 12 October 2007

[3] András Gyenis, Agustin Di Paolo, Jens Koch, Alexandre Blais, Andrew A. Houck, and David I. Schuster, Moving beyond the Transmon: Noise-Protected Superconducting Quantum Circuits, PRX Quantum **2**, 030101 – Published 2 September 2021.

[4] A. Salmanogli, "Entanglement Engineering by Transmon Qubit in a Circuit QED," *arXiv preprint arXiv:2109.00316*, 2021.

[5] Alexandre Blais, Arne L. Grimsmo, S. M. Girvin, and Andreas Wallraff, Circuit quantum electrodynamics, Rev. Mod. Phys. **93**, 025005 – Published 19 May 2021.

[6] P. Magnard, S. Storz, P. Kurpiers, J. Schär, F. Marxer, J. Lütolf, T. Walter, J.-C. Besse, M. Gabureac, K. Reuer, A. Akin, B. Royer, A. Blais, and A. Wallraff1, Microwave Quantum Link between Superconducting Circuits Housed in Spatially Separated Cryogenic Systems, PHYSICAL REVIEW LETTERS 125, 260502 (2020).

[7] Marie Lu, Jean-Loup Ville, Joachim Cohen, Alexandru Petrescu, Sydney Schreppler, Larry Chen, Christian Jünger, Chiara Pelletti, Alexei Marchenkov, Archan Banerjee, William P. Livingston, John Mark Kreikebaum, David I. Santiago, Alexandre Blais, and Irfan Siddiqi, Multipartite Entanglement in Rabi-Driven Superconducting Qubits, PRX Quantum **3**, 040322 – Published 23 November 2022.

[8] A Salmanogli, Entangled state engineering in the 4-coupled qubits system, Physics Letters A 479, 128925, 2023.

[9] A. Salmanogli, Qubit coupling to reservoir modes: engineering the circuitry to enhance the coherence time, arXiv preprint arXiv:2205.13361, 2024.

[10] B. Prabowo, G. Zheng, M. Mehrpoo, B. Patra, P. Harvey-Collard, J. Dijkema, A. Sammak, G. Scappucci, E. Charbon, F. Sebastiano, L. M. K. Vandersypen, M. Babaie, "A 6-to-8GHz 0.17mW/Qubit Cryo-CMOS Receiver for Multiple Spin Qubit Readout in 40nm CMOS Technology", ISSCC 2021 / SESSION 13 / CRYO-CMOS FOR QUANTUM COMPUTING / 13.3, 2021.

[11] J. C. Bardin et al., "A 28nm Bulk-CMOS 4-to-8GHz 2mW Cryogenic Pulse Modulator for Scalable Quantum Computing," 2019 IEEE International Solid-State Circuits Conference - (ISSCC), San Francisco, CA, USA, 2019, pp. 456-458, doi: 10.1109/ISSCC.2019.8662480.

[12] B. Patra et al., "A Scalable Cryo-CMOS 2-to-20GHz Digitally Intensive Controller for 4×32 Frequency Multiplexed Spin Qubits/Transmons in 22nm FinFET Technology for Quantum Computers," 2020 IEEE International Solid-State Circuits Conference - (ISSCC), San Francisco, CA, USA, 2020, pp. 304-306, doi: 10.1109/ISSCC19947.2020.9063109.

[13] L. L. Guevel et al., "A 110mK 295µW 28nm FDSOI CMOS Quantum Integrated Circuit with a 2.8GHz Excitation and nA Current Sensing of an On-Chip Double Quantum Dot," 2020 IEEE International Solid-State Circuits Conference - (ISSCC), San Francisco, CA, USA, 2020, pp. 306-308, doi: 10.1109/ISSCC19947.2020.9063090.

[14] J. -S. Park et al., "A Fully Integrated Cryo-CMOS SoC for Qubit Control in Quantum Computers Capable of State Manipulation, Readout and High-Speed Gate Pulsing of Spin Qubits in Intel 22nm FFL FinFET Technology," 2021 IEEE International Solid-State Circuits Conference (ISSCC), San Francisco, CA, USA, 2021, pp. 208-210, doi: 10.1109/ISSCC42613.2021.9365762.

[15] B. Prabowo et al., "A 6-to-8GHz 0.17mW/Qubit Cryo-CMOS Receiver for Multiple Spin Qubit Readout in 40nm CMOS Technology," 2021 IEEE International Solid-State Circuits Conference (ISSCC), San Francisco, CA, USA, 2021, pp. 212-214, doi: 10.1109/ISSCC42613.2021.9365848.

[16] M. Mehrpoo, B. Patra, J. Gong, P. A. 't Hart, J. P. G. van Dijk, H. Homulle, G. Kiene, A. Vladimirescu, F. Sebastiano, E. Charbon, M. Babaie, "Benefits and Challenges of Designing Cryogenic CMOS RF Circuits for Quantum Computers", 978-1-7281-0397-6/19/$31.00 ©2019 IEEE, 2019.

[17] E. Charbon, "Cryo-CMOS Electronics for Quantum Computing Applications", 978-1-7281-1539-9/19/$31.00 ©2019 IEEE, 2019.

[18] S. Pellerano, S. Subramanian, J-S Park, B. Patra, T. Mladenov, X. Xue, L. M. K. Vandersypen, M. Babaie, E. Charbon, F. Sebastiano, "Cryogenic CMOS for Qubit Control and Readout", IEEE CICC 2022.

[19] A. Ruffino, Y. Peng, T. -Y. Yang, J. Michniewicz, M. F. Gonzalez-Zalba and E. Charbon, "13.2 A Fully-Integrated 40-nm 5-6.5 GHz Cryo-CMOS System-on-Chip with I/Q Receiver and Frequency Synthesizer for Scalable Multiplexed Readout of Quantum Dots," 2021 IEEE International Solid-State Circuits Conference (ISSCC), San Francisco, CA, USA, 2021, pp. 210-212, doi: 10.1109/ISSCC42613.2021.9365758.

[20] U. Alakusu et al., "A 210-284-GHz I-Q Receiver With On-Chip VCO and Divider Chain," IEEE Microwave and Wireless Components Letters, vol. 30, no. 1, pp. 50-53, Jan. 2020.

[21] K. Kang et al., "A Cryo-CMOS Controller IC With Fully Integrated Frequency Generators for Superconducting Qubits," 2022 IEEE International Solid-State Circuits Conference (ISSCC), San Francisco, CA, USA, 2022, pp. 362-364, doi: 10.1109/ISSCC42614.2022.9731574.

[22] Guilliam Butseraen, Arpit Ranadive, Nicolas Aparicio, Kazi Rafsanjani Amin, Abhishek Juyal, et al.. A gate-tunable graphene Josephson parametric amplifier. Nature Nanotech., 2022, 17, pp.1153-1158. ff10.1038/s41565-022-01235-9ff. ffhal-03650619f

[23] V. V. Sivak, S. Shankar, G. Liu, J. Aumentado, and M. H. Devoret, Josephson Array Mode Parametric Amplifier, arXiv:1909.08005v2 [quant-ph] 16 Feb 2022.

[24] A. Salmanogli, A. Bermak, Quantum Parametric Amplification and NonClassical Correlations due to 45 nm nMOS Circuitry Effect, arXiv preprint arXiv:2310.16385, 2023.

[25] M. O. Scully, M. S. Zubairy, Quantum Optics, Cambridge University Press, UK, 1997.

[26] J. R. Johansson, P. D. Nation, and F. Nori, QuTiP 2: A Python framework for the dynamics of open quantum systems, 2013 Comp. Phys. Comm. 184, 1234.

[27] A. salmanogli, Quantum correlation of microwave two-mode squeezed state generated by nonlinearity of InP HEMT, Scientific Reports 13 (1), 11528, 2023.

[28] A. Salmanogli, Enhancing quantum correlation at zero-IF band by confining the thermally excited photons: InP hemt circuitry effect, Optical and Quantum Electronics 55 (8), 745, 2023.

[29] Schlichtholz, K., Woloncewicz, B. & Żukowski, M. Simplified quantum optical Stokes observables and Bell's theorem. *Sci Rep* **12**, 10101 (2022). https://doi.org/10.1038/s41598-022-14232-8.

[30] Daniel F. V. James , Paul G. Kwiat, William J. Munro and Andrew G. White, On the Measurement of Qubits, Phys. Rev. A **64**, 052312 – Published 16 October 2001.

[31] E. Cha, N. Wadefalk, G. Moschetti, A. Pourkabirian, J. Stenarson and J. Grahn , "InP HEMTs for Sub-mW Cryogenic Low-Noise Amplifiers", IEEE Electron Device Letters, vol. 41, no. 7, pp. 1005-1008, 2020, doi: 10.1109/LED.2020.3000071.

[32] E. Cha, N. Wadefalk, P. Nilsson, J. Schleeh, G. Moschetti, A. Pourkabirian, S. Tuzi, J. Grahn, "0.3–14 and 16–28 GHz Wide-Bandwidth Cryogenic MMIC Low-Noise Amplifiers", IEEE Transactions on Microwave Theory and Techniques, vol. 66, no. 11, pp. 4860-4869, Nov. 2018, doi: 10.1109/TMTT.2018.2872566.

[33] A Salmanogli, Squeezed state generation using cryogenic InP HEMT nonlinearity, Journal of Semiconductors 44 (5), 052901.

[34] Cramer, M., Plenio, M., Flammia, S. *et al.* Efficient quantum state tomography. *Nat Commun* **1**, 149 (2010). https://doi.org/10.1038/ncomms1147.

[35] E. Toninelli, B. Ndagano, A. Vallés, B. Sephton, I. Nape, A. Ambrosio, F. Capasso, M. Padgett, and A. Forbes, "Concepts in quantum state tomography and classical implementation with intense light: a tutorial," Adv. Opt. Photon. 11, 67-134 (2019).

[36] Cohadon, PF. Improved squeezing of noise. *Nat. Photonics* **14**, 202–204 (2020). https://doi.org/10.1038/s41566-020-0616-y

[37] Adriano A Batista, Squeezing of thermal noise in a parametrically-driven oscillator, 2011 *J. Phys.: Conf. Ser.* **285** 012041.

[38] W. Lee and E. Afshari, "An 8GHz, 0.45dB NF CMOS LNA employing noise squeezing," 2011 IEEE Radio Frequency Integrated Circuits Symposium, Baltimore, MD, USA, 2011, pp. 1-4, doi: 10.1109/RFIC.2011.5940695.

[39] W. Lee and E. Afshari, "A CMOS Noise-Squeezing Amplifier," in IEEE Transactions on Microwave Theory and Techniques, vol. 60, no. 2, pp. 329-339, Feb. 2012, doi: 10.1109/TMTT.2011.2178318.

[40] A. Salmanogli, D. Gokcen and H. S. Gecim, "Entanglement Sustainability in Quantum Radar," in IEEE Journal of Selected Topics in Quantum Electronics, vol. 26, no. 6, pp. 1-11, Nov.-Dec. 2020, Art no. 9200211, doi: 10.1109/JSTQE.2020.3020620.

[41] A. Salmanogli and H. S. Geçim, "Optical and Microcavity Modes Entanglement by Means of Plasmonic Opto-Mechanical System," in IEEE Journal of Selected Topics in Quantum Electronics, vol. 26, no. 3, pp. 1-10, May-June 2020, Art no. 4600110, doi: 10.1109/JSTQE.2020.2987171.

[42] A. Salmanogli, Quantum analysis of plasmonic coupling between quantum dots and nanoparticles, Phys. Rev. A **94**, 043819 – Published 13 October 2016

[43] Ahmad Salmanogli, Dincer Gokcen, and H. Selcuk Gecim, Entanglement of Optical and Microcavity Modes by Means of an Optoelectronic System, Phys. Rev. Applied **11**, 024075 – Published 28 February 2019.

[44] A. Salmanogli, Modification of a plasmonic nanoparticle lifetime by coupled quantum dots, Phys. Rev. A **100**, 013817 – Published 10 July 2019.